\documentclass[prl,twocolumn,superscriptaddress,preprintnumbers,amsmath,amssymb]{revtex4}
\usepackage{graphicx}
\usepackage{amsmath}
\usepackage{amssymb}
\usepackage{epsfig}
\usepackage{wasysym}
\usepackage{bbm}
\renewcommand{\narrowtext}{\begin{multicols}{2} \global\columnwidth20.5pc}
\renewcommand{\v}[1]{{\bf #1}}

\def\be{\begin{eqnarray}}
\def\ee{\end{eqnarray}}

\newcommand{\Eq}[1]{Eq.~(\ref{#1})}

\newcommand \im{i}
\newcommand \dif{{\rm d}}

\newcommand	\vecQ{\vec{Q}}
\newcommand \veck{\vec{k}}

\newcommand{\nd}{{\vphantom{\dagger}}}

\begin{document}
\draft

\title{Antiferromagnetic Correlation and the Pairing Mechanism of the Cuprates and Iron Pnictides : a View From the Functional Renormalization Group
Studies}

\author{Fa Wang}
\author{Hui Zhai}
\author{Dung-Hai Lee}

\affiliation{Department of Physics,University of California at Berkeley,
Berkeley, CA 94720, USA}
\affiliation{Materials Sciences Division,
Lawrence Berkeley National Laboratory, Berkeley, CA 94720, USA.}

\begin{abstract}

We compare the one-loop functional renormalization group results for the cuprates and the iron pnictides.
Interestingly a coherent picture suggesting that antiferromagnetic correlation causes pairing
for both materials emerges.

\end{abstract}

\date{\today}
\maketitle

In the search for high temperature superconductors two classes of materials with $T_c$ above 50 K have been found - the cuprates\cite{Cu} and the iron pnictides\cite{Fe}. There are striking similarities between them: (i) both have layered structure, (ii)  the parent (non-superconducting) compound for both exhibit antiferromagnetic (AF) order (although with different ordering wavevector), (iii) both materials become superconducting (SC) upon doping. On the other hand, there are also important differences. (i) The parent compounds of the cuprates are Mott insulators while those for the iron pnictides are metallic. (ii) The cuprates are effectively one-band\cite{zhang-rice} materials while the iron-pnictides have multi bands at the Fermi energy\cite{FS-iron,kuroki}. (iii) The gap function of the cuprates has $d_{x^2-y^2}$ symmetry\cite{d-wave}, hence has nodes, while current evidences suggest that the iron pnictides has $s$-wave pairing symmetry \cite{no-node}. In addition to the above, there is another similarity between the cuprates and the iron pnictides - there is no consensus on the pairing mechanism.

Numerous attempts have been made to uncover the pairing mechanism of the cuprates. It is reasonable to expect considerable efforts will be devoted to that of the iron pnictides as well. In this paper we compare the one-loop functional renormalization group (1LFRG) results for both materials\cite{Honerkamp,Fa}. Interestingly, a coherent picture pointing to the involvement of the antiferromagnetic correlation in the superconducting pairing emerges.

Currently there is a lack of an ideal first-principle approach for strongly correlated systems. For example, direct diagonalization and density matrix renormalization group are limited by the small system size. Monte-Carlo simulation is hindered by the fermion sign problem. Mean-field and variational wavefunction approach are not unbiased. The 1LFRG method used to gain the results in this paper is unbiased\cite{shankar}, and can be applied to infinite systems for a range of interaction strength. However it is not a systematic expansion of a small parameter. Applying this method to the cuprates, Honerkamp {\it et.al.} found that effective interaction favoring both the AF order and $d_{x^2-y^2}$ pairing were generated\cite{Honerkamp}. Recently we generalized this method and applied it to iron pnictides. Interestingly, effective interaction favoring the $(\pi,0)/(0,\pi)$ AF order, and an extended s-wave pairing with opposite sign in electron and hole packet \cite{Fa} appears. In this paper we compare the results of 1LFRG for these two materials in hoping for hints for the paring mechanism for both. However, we would like to emphasize that when applying the 1LFRG to the cuprates our best hope is to describe the high temperature superconductors on the overdoped side.

\begin{figure}[tbp]
\includegraphics[scale=.38]{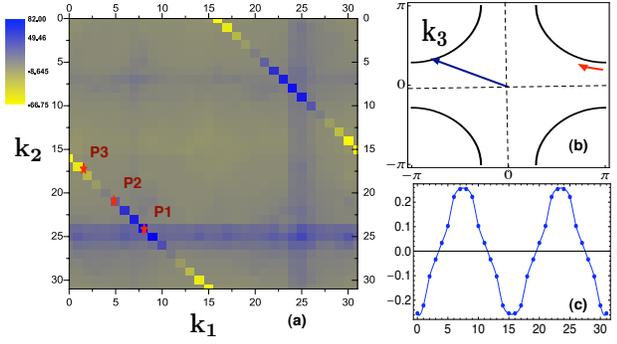}
\caption{(a) The renormalized $V({\bf k_1},{\bf k_2},{\bf k_3},{\bf k_4})$ for the one-band Hubbard model used in Ref.\cite{Honerkamp}. The momentum $\v k_3$ is fixed at the position shown in part (b), and the scattering amplitude is plotted as  ${\bf k_1}$ and ${\bf k_2}$ go around the Fermi surface started from the place indicated by the small red arrow. In constructing this plot the Fermi surface is discretized into $32$ parches.  (b) The Fermi surface of the single band Hubbard model in Ref.\cite{Honerkamp}.  (c) The gap function deduced from the effective pair scattering amplitudes is plotted as a function of momentum on the discretized Fermi surface. 
Here we choose the parameter as $t=1eV$, $t^\prime=0.3eV$ and $U=3eV$}\label{vertex-1}
\end{figure}

\begin{figure}[bp]
\includegraphics[scale=.38]{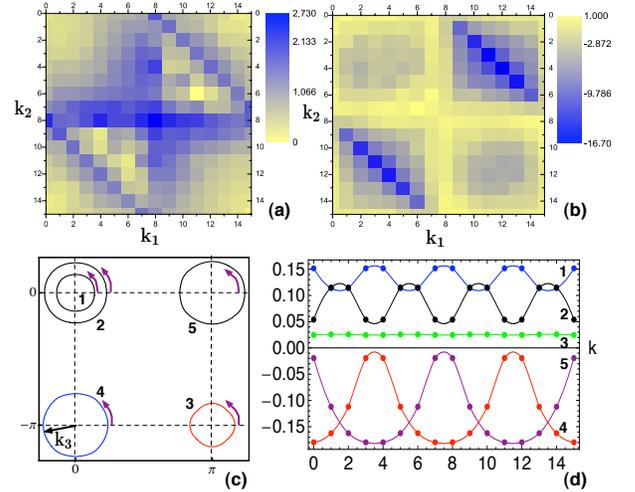}
\caption{(a) The renormalized umklapp scattering amplitudes $V({\bf k_1},{\bf k_2},{\bf k_3},{\bf k_4})$ as two electrons are scattered from the small hole pocket around $(\pi,-\pi)$ (the red circle in panel (c)) to the electron pocket (the blue circle in panel (c)) around $(0,-\pi)$.  
The momentum $\v k_3$ is fixed on the electron pocket as shown in panel (c), and the scattering amplitudes is plotted as ${\bf k_1}$ and ${\bf k_2}$ go around the hole Fermi surface around $(\pi,-\pi)$, started from the placed indicated by the small purple arrow. Here ${\bf k_4}$ is chosen to lie in the same band as ${\bf k_3}$. (b) The renormalized intra-band scattering amplitudes $V({\bf k_1},{\bf k_2},{\bf k_3},{\bf k_4})$. Here all momenta lie ont the electron Fermi surface around $(0,-\pi)$.  In constructing panel (a) and (b) each Fermi surface is discretized into $16$ parches. (c) The Fermi surfaces for the five band model in Ref.\cite{Fa} with the position of $\v k_3$ marked. (d) The gap function deduced from the effective pair scattering amplitudes as the momentum goes around the five Fermi surfaces, started from the places indicated by the small purple arrows. Here the interaction parameters are chosen as $U_1=4eV$, $U_2=2eV$, $J_H=0.7eV$, and doping is $0.05$ hole doped.
}\label{vertex-5}
\end{figure}

The goal of the 1LFRG is to generate an effective two-particle scattering
\be V(\v 1,\v 2;\v 3,\v 4)\psi^\dag_{\v 3 s}\psi^\dag_{\v 4 s^\prime}\psi_{\v 2 s^\prime}\psi_{\v 1 s},\label{eff}\ee
 where $\v 1,..,\v 4$  each stands for momentum and band index, and $s,s'$ are spin labels. 
 In Fig.\ref{vertex-1}(a) and Fig. \ref{vertex-5}(a), we fix $\v 3$ and plot $V(\v 1,\v 2,\v 3,\v 4)$ as $\v 1$ and $\v 2$ run around the Fermi surface. (Once $\v 1,\v 2,\v 3$ are given, $\v 4$ is determined (see below).) Fig. \ref{vertex-1}(a) is for the single band Hubbard model as applied to the cuprates\cite{Honerkamp}, which is written as
\begin{equation}
H=-t\sum\limits_{\langle ij\rangle\sigma}c^\dag_{i\sigma}c_{j\sigma}+t^\prime\sum\limits_{\langle\langle ij\rangle\rangle}c^\dag_{i\sigma}c_{j\sigma}+U\sum\limits_{i}n_{i\uparrow}n_{i\downarrow}
\end{equation} 
where $\langle ij\rangle$ is nearest neighboring sites and $\langle\langle ij\rangle\rangle$ are next-nearest sites. Since there is only one band, $\v 1,..,\v 4$ are just wavevectors. Here once $\v k_1,\v k_2$ and $\v k_3$ are given, $\v k_4$ is fixed by momentum conservation. Fig. \ref{vertex-5}(a) is for the iron pnictides. Here we use a five band mode written as
\begin{eqnarray}
&&H=\sum_{\v{k},s}\sum_{a,b=1}^5 c_{a\v{k} s}^\dagger K_{ab}(\v{k})c_{b\v{k} s}+\sum_{i}\Big\{U_1\sum_{a}
 n_{i,a,\uparrow}n_{i,a,\downarrow}\nonumber\\&&+U_2\sum_{a< b}n_{i,a} n_{i,b}+
 J_H\sum_{a< b,ss^\prime}c_{ia s}^\dagger c_{ib s'}^\dagger c_{ia s'}^\nd c_{ib s}\nonumber\\&&+J_H\sum\limits_{a<b} (c^\dagger_{ia\uparrow}c^\dagger_{ia\downarrow}c_{ib\downarrow}c_{ib\uparrow}+h.c.) \Big\}.
\end{eqnarray}
The parameters used in constructing  $K_{ab}(\v k)$ can be found in Ref.~\cite{kuroki}. Here $\v 1=(\v k_1, a), \v 2=(\v k_2,a)$ and $\v 3=(\v k_3,b), \v 4=(\v k_4,b)$ where $a, b$ labels the band that produces the hole-like Fermi surface around $(\pi,\pi)$ or the band which produces the electron-like Fermi surface around the $(0,\pi)$ point. Again, $\v k_4$ can be determined from  $\v k_1,\v k_2,\v k_3$ by momentum conservation.

There are two main features in these plots. First, the blue vertical and horizontal stripes in Fig.1(a) and 2(a)  indicate {\it strong positive} scattering amplitudes. The momenta in the horizontal stripe satisfy ${\bf k_2}={\bf k_3}+{\bf Q}$, and $\v k_1={\bf k_4}-\v Q$, while those in the vertical stripe satisfy ${\bf k_1}={\bf k_3}+{\bf Q}$, and $\v k_2={\bf k_4}-{\bf Q}$. In the above ${\bf Q}\approx (\pi,\pi)$ for cuprates (Fig.1(a)) and ${\bf Q}\approx (\pi,0)$ for iron pnictides (Fig.2(a)). Each of the scattering process
$V(\v 1,\v 2;\v 3,\v 4)\psi^\dag_{\v 3 s}\psi^\dag_{\v 4 s^\prime}\psi_{\v 2 s^\prime}\psi_{\v 1 s}$
in the horizontal stripe has a corresponding process
$V(\v 1,\v 2;\v 4,\v 3)\psi^\dag_{\v 4 s}\psi^\dag_{\v 3 s^\prime}\psi_{\v 2 s^\prime}\psi_{\v 1 s}$
 in the vertical one. The fact that the amplitudes associated with both are strong implies that if we
 decompose \Eq{eff} into the sum of singlet and triplet channels
 \be
 V_s(\v 1,\v 2;\v 3,\v 4)(\psi^\dag_{\v 3 s}\psi^\dag_{\v 4 s^\prime}-\psi^\dag_{\v 3 s^\prime}\psi^\dag_{\v 4 s})(\psi_{\v 2 s^\prime}\psi_{\v 1 s}-\psi_{\v 2 s}\psi_{\v 1 s^\prime})\nonumber\ee
 and
 \be V_t(\v 1,\v 2;\v 3,\v 4)(\psi^\dag_{\v 3 s}\psi^\dag_{\v 4 s^\prime}+\psi^\dag_{\v 3 s^\prime}\psi^\dag_{\v 4 s})(\psi_{\v 2 s^\prime}\psi_{\v 1 s}+\psi_{\v 2 s}\psi_{\v 1 s^\prime}),\nonumber\ee
 it is $V_s$ that dominates the scattering amplitude of the horizontal (and vertical) stripe. This is not all that surprising for systems with strong short-range interactions.
Having a positive amplitude, the scattering associated with the horizontal stripe tends to drive the AF order. The scattering associated with the vertical stripe would drive charge density wave (CDW) order had the amplitude been negative. With the wrong sign, as in Fig.1(a) and 2(a), CDW is not favored. 

The second notable feature of Fig.\ref{vertex-1}(a) and \ref{vertex-5}(a,b) is the diagonal stripes. The momenta in these stripes satisfy $\v k_1+\v k_2=0$ hence the corresponding process are Cooper scattering. While the color of the diagonal stripe changes (which represents sign change in the scattering amplitude) in Fig.\ref{vertex-1}(a), it stays the same in Fig.\ref{vertex-5}(a,b). 
The sign changes in Fig.1(a) implies that the effective pairing interaction $V_{\rm pairing}(\v k,\v k')$ changes sign four times as $\v k$ moves around the Fermi surface with $\v k'$ fixed. The latter is the hallmark of the $d_{x^2-y^2}$ pairing symmetry. The diagonal stripes in Fig.\ref{vertex-5}(a) and (b) are associated with the {\it inter-band} and {\it intra-band} pair scattering
respectively. The fact that the inter-band pair scattering amplitudes are positive (Fig.2(a)) does not mean they disfavor pairing. Because a wrong (positive) sign in the inter-band Cooper scattering can always be absorbed by making the sign of the gap function opposite on the two Fermi surfaces\cite{twoband}.

The fact that the horizontal SDW (and the associated vertical) stripe intersects the {\it inter-band} rather than {\it intra-band} diagonal (Cooper scattering) stripes is responsible for the difference in pairing symmetry between the cuprates and iron pnictides.
To understand that we first note that while an uniform positive inter-band pair scattering drives pairing, intra-band pairing requires the presence of negative pair scattering. Secondly, the scattering processes associated with the intersection of the horizontal and diagonal stripes, namely,
\be 
\psi^\dag_{{\bf -k_1}-{\bf Q},s}\psi^\dag_{{\bf k_1}+{\bf Q},s^\prime}\psi_{{\bf -k_1},s^\prime}\psi_{{\bf k_1},s}\label{afsc}\ee drive both AF and SC. Indeed, SC and AF appear as different decoupling of \Eq{afsc}, with \be \langle\psi^\dag_{{\bf -k_1}-{\bf Q},s}\psi^\dag_{{\bf k_1}+{\bf Q},s^\prime}\rangle\ne 0, ~~ \langle\psi_{{\bf -k_1},s^\prime}\psi_{{\bf k_1},s}\rangle\ne 0\nonumber\ee describing SC, and  \be \langle\psi^\dag_{{\bf -k_1}-{\bf Q},s}\psi_{{\bf -k_1},s^\prime}\rangle\ne 0,~~\langle\psi^\dag_{{\bf k_1}+{\bf Q},s^\prime}\psi_{{\bf k_1},s}\rangle\ne 0\nonumber\ee describing AF. Because AF correlation requires the scattering amplitudes to be positive, the sign of the Cooper scattering corresponding to the intersection is fixed (to be positive). Under this constraint, the only way that overall pairing can be favored for the one-band case (Fig.\ref{vertex-1}(a)) is for the pairing interaction to change sign. Since the diagonal stripes intersect both the vertical and horizontal stripes, the pairing interaction $V_{\rm pairing}(\v k,\v k')$ is forced to change sign four times as $\v k$ moves around the Fermi surface. As discussed earlier, this leads to the $d_{x^2-y^2}$ pairing. For iron pnictides the intersection corresponds to inter-band rather than intra-band pair scattering. Here there is no problem for all inter-band pair scattering amplitudes to stay positive; all that is required is for the gap function to take on opposite sign on the electron and hole Fermi surfaces. In this way, antiferromagnetic correlation naturally leads to an out-of-phase s-wave pairing.

\begin{figure}[tbp]
\includegraphics[scale=.39]{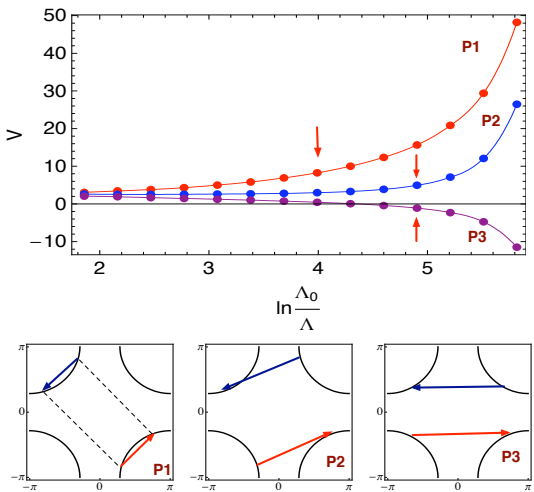}
\caption{(a) The renormalization group flow of three different types of pair scattering amplitude for the single-band Hubbard model in Ref.\cite{Honerkamp}. $P_1$ is the scattering process that drives both SDW and pairing. $P_2$ and $P_3$ are two other pair scattering processes which are not in the SDW channel (see Fig.1(a)). The red arrows mark the renormalization group steps at which the scattering amplitudes begin to increase rapidly.  (b) Schematic representation of $P_{1,2,3}$. The red arrow represents ${\bf k_4-k_2}$ and the blue arrow denotes ${\bf k_3-k_1}$.  }\label{RG-1}
\end{figure}

\begin{figure}[tbp]
\includegraphics[scale=.4]{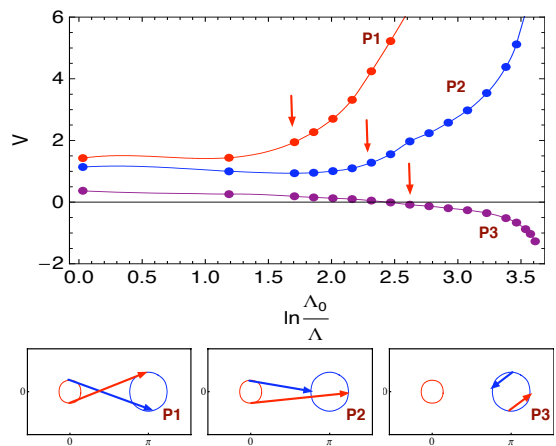}
\caption{(a) The renormalization group flow of three different types of pair scattering amplitude for the five-band Hubbard-Hunds model in Ref.\cite{Fa}.  $P_1$ is the scattering process that drives both SDW and pairing. $P_2$ is a generic pair scattering processes which is not in the SDW channel. $P_3$ is an intra-packet pair scattering process. The red arrows mark the renormalization group steps at which the scattering amplitudes begin to increase rapidly.  (b) Schematic representation of $P_{1,2,3}$. The red arrow represents ${\bf k_4-k_2}$ and the blue arrow denotes ${\bf k_3-k_1}$.  }\label{RG-5}
\end{figure}

In the following we shall provide more numerical evidence that the superconducting pairing is driven by the antiferromagnetic correlation. This is achieved by monitoring the growth of the SDW and pairing interaction during the RG. In Fig.\ref{RG-1}(a) and \ref{RG-5}(a) we plot the RG flow of three pair scattering processes labeled as $P_1,P_2,P_3$. Here $P_1$ is at the intersection of the horizontal and diagonal stripes. This is the type of interaction that has the dual characteristics of being both SDW scattering and pairing interaction as discussed earlier. For Fig.\ref{RG-1}(a) $P_{2,3}$ are two other generic  pairing interaction. For Fig.\ref{RG-5}(a) $P_2$ is a generic inter-band pair scattering while $P_3$ is an intra-band Cooper scattering. As one can see, in both figures the $P_1$ process (in fact the processes associated with the entire horizontal stripe) grows first. When $P_1$ gets strong, the magnitude of the other generic pairing interaction ($P_2$ and $P_3$) grows. This suggest that it is the AF correlation (i.e., strong SDW scattering) that drives SC! 

\begin{figure}[tbp]
\includegraphics[scale=.45]{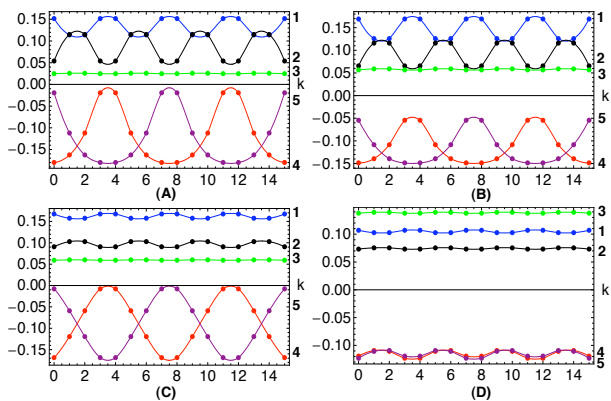}
\caption{The gap function of iron pinicide as momentum goes around five Fermi surfaces for different interaction and doping parameters  (A), $J_{H}=0.7eV$, $0.05$ hole hoped; (B):  $J_{H}=0.1eV$, $0.05$ hole hoped; (C): $J_{H}=0.7eV$, $0.10$ electron hoped; (D) $J_{H}=0.3eV$, $0.10$ electron hoped. For all of four plots $U_1=4eV$, $U_2=2eV$.  }\label{Formfactor}
\end{figure}

As shown in Fig.2(d) the gap function of iron pnictides is quite anisotropic on the electron Fermi surface (the anisotropy is smaller on the hole Fermi surfaces.) Our results suggest that the degree of such anisotropy depends on the interaction parameters 
as well as doping, as shown in Fig. \ref{Formfactor}. In the extreme case, the gap function can even change sign (hence exhibit nodes) on the electron Fermi surfaces. 

We would like to point out the fact \Eq{afsc} can be ``decouple'' in both the antiferromagnetic and the superconducting channels is reminiscent of the spirit of the so-called pairing decoupling of the AF exchange interaction in the ``RVB'' theory of the cuprates\cite{baskaran}. Similar pairing decoupling of the AF exchange has been made in Ref.\cite{Hu,Fuchun} for the iron pnictides. However, we should stress that the final effective interaction generated by the our FRG is {\it not} a simple spin-spin exchange interaction as that described in Ref.\cite{Hu}.

Before closing we would like to propose an experiment which can in principle detect the signature of the out-of-phase s-wave pairing discussed in this paper. The idea is to study the quasiparticle interference\cite{wanglee} using STM\cite{hoffman}. If the electron pocket and hole pocket have out-of-phase order parameter, the Nambu spinor associated with the quasiparticle at the electron and hole Fermi surfaces will be orthogonal. (Note that from the angle-resolved photoemission\cite{hong} the gap value for the electron pocket is almost identical to that of the larger-gap hole-pocket. As a result, scattering from electron to hole pocket is an allowed elastic process in the superconducting state.) For example, under the gauge where the order parameter is real, one of them will be $\sim \begin{pmatrix}1\cr 1\end{pmatrix}$, and the other $\sim \begin{pmatrix}1\cr -1\end{pmatrix}$. As a result, a scalar impurity (which operates as $\begin{pmatrix}1&0\cr 0&-1\end{pmatrix}$ in the Nambu space) can not scatter quasiparticle between two electron pockets, while can do so between the electron and hole pockets. As a result, for bias at the larger gap  edge ($\sim 12 meV$), the interference peaks (rings) surrounding the reciprocal lattice vector (here we use the unit cell containing two Fe atoms) will be absent in the Fourier transformed STM spectroscopy. In contrast the peaks surrounding $(\pm \pi,\pm\pi)$ will be present. The missing peaks around the reciprocal lattice vector will recover as the bias is increases from the gap edge. If such behavior is seen this is an evidence of the our-of-phase s-wave pairing.

In summary, we have shown that within the one-loop functional renormalization group approach the pairing in both the cuprates and iron pnictides are driven by the antiferromagnetic correlation. We have shown that this naturally leads to d-wave pairing symmetry for the cuprates, where magnetic fluctuation is intra-band, and an out-of-phase s-wave pairing symmetry for the iron pnictides, where magnetic fluctuation is inter-band. Finally, in addition to these two family of compounds there are other instances of superconductivity occurring upon exiting the  antiferromagnetic phase (examples include the heavy fermion and the organic compounds\cite{others}). It is possible that the mechanism discussed in the present paper is applicable to those as well.

\end{document}